\documentclass[onecolumn,preprintnumbers,superscriptaddress,amsmath,amssymb,aps]{revtex4}
\usepackage{graphicx,float} 
\usepackage{tensor}
\usepackage{mathrsfs}
\usepackage{amsfonts}
\usepackage{lipsum}
\usepackage{caption}

\newcommand{\Lag}{\mathscr{L}}

\begin{document}

\title{Is Parity Violation a Dynamical Effect?}
\author{James H. Atwater, David Lambert, Yuri Rostovtsev}
\affiliation{{Center for Nonlinear Sciences and Department of Physics, University of North Texas},{Denton},{TX}, {76203}, {USA}}
\date{May 2026}

\begin{abstract}    
As has been shown by multiple authors in recent decades, it is possible to reformulate various portions of the standard model over the ring of complex quaternions. In this paper, we utilize a complex quaternion spin representation of the spacetime algebra to derive the magnetic moments of standard model fermions and the $W^\pm$ boson. The moments calculated are not limited to those with the photon. We account for coupling to the magnetic fields of each standard model gauge boson. We naively assume that fermions coupled to weak isospin have magnetic moments with the charged bosons. Upon assuming these charged moments exist, we realize that they are to be influenced by the neutral, pseudovector-valued magnetic fields that are observed when a charged particle is moving. Visualizing the derived moments of fermions and charged bosons together, we find a possible explanation for the parity asymmetry observed in charged weak interactions.
\end{abstract}

\maketitle

\section{Introduction}

Since the observations made in the Stern-Gerlach experiment~\cite{GerlachStern1922,FriedrichHerschbach2003}, magnetic moments~\cite{Skipper2025,ALIBERTI20251} have been among the most crucial quantities related to the establishment and verification of theories of particles. Its importance is obvious as the magnetic moment of a particle encodes its charge, mass, as well as information about how its spin angular momentum couples to an external magnetic field. As quantum field theory superseded quantum mechanics, magnetic moments maintained their significance, and to this day, the anamolous magnetic moment of the electron remains the crown jewel of theoretical accuracy. As we began to piece together what is now the standard model of particle physics, we were confronted with the reality that there are more magnetic fields than the photon's alone with which fundamental particles may have a moment.

This reality is a simple consequence of the fact that there is more than one gauge boson mediating interactions between fundamental particles. Each of these gauge bosons has a momentum (field strength) whose purely spatial components are the components of the corresponding magnetic field. Not all gauge bosons are equal however, and in the electroweak sector post-spontaneous symmetry breaking this is most clear. We have a neutral massless boson - the photon, a neutral massive boson - the $Z$ boson, and a charged massive boson - the $W^\pm$. The mass of the latter two bosons severely limits their range, as Yukawa figured out long before we identified them \cite{Yukawa1934}.

In an attempt to better understand these exotic magnetic fields, we take the non-relativistic limit of the standard model and, in the process, identify explicitly the moments of each lepton, quark, and charged boson with any magnetic field able to exert a force on them. Upon doing so, we realize the non-relativistic limit conceals certain details. 

One staggering result of organizing the standard model was the confirmation of parity violation in charged weak decays. Predicted by Yang and Lee in \cite{LeeYang1956} and observed by Wu in ~\cite{Wu1957}, this novel asymmetry forced us to reconcile that the universe may not be ambidextrous. Today, the parity violating nature of these interactions is hard-wired into the standard model and manifests itself in the form of vector minus axial vector couplings. However, no satisfactory explanation or motivation for this asymmetry has been established. Our experience with this violation has left us with the demoralizing impression that the universe can somehow tell the difference between left and right. We propose here that the universe does not know one direction from the other, but magnetic moments do.

The hypercomplex approach to particle physics we take draws inspiration from the work of \cite{Dirac1944,Hestenes1973,Rotelli1989,DeLeoRotelli1995,DeLeoRotelli1996,KravchenkoShapiro1996,DeLeo1997,DeLeo2001,Morita1986,Morita2007,Furey2015,Kravchenko2003}. The calculations we present utilize a complex quaternion spin representation of the spacetime algebra that was shown to function equivalently to the Dirac representation in \cite{Atwater2026}.

\section{Methods of Calculation}

Consider four $2\times2 $ complex quaternion matrices labeled by a spacetime index $\mu=0,1,2,3$ $$\Gamma^\mu = \left( \begin{pmatrix}
1 & 0 \\
0 & -1 
\end{pmatrix} , 
\begin{pmatrix}
0 & i\omega_\ell \\
-i\omega_\ell & 0 
\end{pmatrix} \right)^T=\left( \begin{pmatrix}
1 & 0 \\
0 & -1 
\end{pmatrix} , 
\begin{pmatrix}
0 & \Sigma_\ell \\
\tilde\Sigma_\ell & 0 
\end{pmatrix} \right)^T,$$
where $\omega_\ell=(h,j,k)$ for $\ell=1,2,3$ is a 3-vector of quaternions and $i$ is a complex number that commutes with each quaternion. These matrices generate a spin representation of the spacetime Clifford algebra since $\{\Gamma^\mu,\Gamma^\nu\}=2\eta^{\mu\nu}$, and for all intents and purposes these matrices are functionally identical to the Dirac representation because $\tilde\Sigma_\ell$ and $\Sigma_\ell$ are functionally identical to the left and right chiral Pauli matrices (usually denoted $\tilde\sigma_\ell$ and $\sigma_\ell$ respectively). The Dirac equation may be written as
$$(i\Gamma^\mu \partial_\mu-m)\psi=0,$$
where the Dirac spinor is defined as $\psi= (\psi_L, \psi_R)^T \in \mathbb{H}^2 \otimes_\mathbb{R} \mathbb{C}.$ The spin representation is a map $S:Cl_{(1,3)}\mathbb{R}\rightarrow \mathfrak{gl}_2\mathbb{H}\otimes\mathbb{C}$ whose image is  $$S(Cl_{(1,3)}\mathbb{R})=\mathbb{R}\{I,\Gamma^\mu,\Gamma^\mu \Gamma^\nu,\Gamma^\mu \Gamma^\nu \Gamma^\lambda,\Gamma^4\}\subseteq \mathfrak{gl}_2\mathbb{H}\otimes\mathbb{C}.$$
The matrix $\Gamma^4=\Gamma^0\Gamma^1\Gamma^2\Gamma^3$ is the pseudoscalar of the Clifford algebra. As with any representation of the spacetime algebra, spin or otherwise, the information of any tensor-valued quantity on spacetime is naturally specified through a homomorphic embedding $$E:S(Cl_{(1,3)}\mathbb{R})\otimes\mathbb{R}^{1,3}\rightarrow S(Cl_{(1,3)}\mathbb{R}) $$determined by the choice of metric, which here is the Minkowski with $\text{diag}(\eta_{\mu\nu})=(+1,-1,-1,-1)$. In the spin representation, the Clifford product is specified by $$\Gamma^\mu \Gamma^\nu=\frac12\{\Gamma^\mu,\Gamma^\nu\}+\frac12[\Gamma^\mu,\Gamma^\nu]=\eta^{\mu\nu}+\frac12[\Gamma^\mu,\Gamma^\nu]$$
The image of an arbitrary vector field $V^\mu=(V_0,\vec{V})^T\in\mathbb{R}^{1,3}$ under $E$ is $$V\equiv E(V^\mu)=\Gamma^\mu\eta_{\mu\nu} V^\nu=\Gamma^\mu V_\mu =\begin{pmatrix}
    V_0 & -i\omega_\ell V_\ell \\
    i\omega_\ell V_\ell & -V_0
\end{pmatrix}.$$
As always, the derivative is defined covariant, $\partial_\mu=(\partial_0, \partial_\ell)$, and we see the appropriate difference in sign to the spatial components of the contravariantly-defined vector field manifest itself here $$d=\Gamma^\mu\partial_\mu= \begin{pmatrix}
    \partial_0 & i\omega_\ell\partial_\ell \\
    -i\omega_\ell\partial_\ell & -\partial_0
\end{pmatrix}.$$
The above expressions for a vector field and the derivative are all that is needed to embed arbitrary tensors and products of them. For instance, the field strength $V_{\mu\nu}=\partial_\mu\wedge V_\nu=\partial_\mu V_\nu-\partial_\nu V_\mu$ of $V^\mu$ is $$E(V_{\mu\nu})=\frac{1}{2}[\Gamma^\mu,\Gamma^\nu]\partial_\mu V_\nu=\begin{pmatrix}
    -\omega_\ell\epsilon^{\ell mn}\partial_mV_n & i\omega_\ell(-\partial_0V_\ell-\partial_\ell V_0) \\
    i\omega_\ell(-\partial_0V_\ell-\partial_\ell V_0) & -\omega_\ell\epsilon^{\ell mn}\partial_mV_n
\end{pmatrix}$$
If the vector field is divergenceless - i.e. if $\partial_\mu V^\mu=\partial_0 V_0+\partial_\ell V^\ell=0$, then $E(\partial_\mu\otimes V_\nu)=E(\partial_\mu V_\nu)=E(\partial_\mu\wedge V_\nu)=E(V_{\mu\nu})= dV$, since $\frac{1}{2}\{\Gamma^\mu,\Gamma^\nu \}\partial_\mu V_\nu=\begin{pmatrix}
    \partial_\mu V^\mu & 0 \\
    0 & \partial_\mu V^\mu
\end{pmatrix}=0$. It is useful to define the components of the field strength as 3-vectors: $\alpha_\ell=-\partial_0 V_\ell-\partial_\ell V_0$, $\beta_\ell=\epsilon_{\ell mn}\partial^mV^n$. In terms of these 3-vector fields, $$F\equiv E(V_{\mu\nu})=\begin{pmatrix}
    -\omega_\ell\beta_\ell & i\omega_\ell\alpha_\ell \\
    i\omega_\ell\alpha_\ell & -\omega_\ell\beta_\ell
\end{pmatrix}.$$
If we take the derivative of the field strength and set it equal to zero, we get a set of four equations equivalent to the source-free Maxwell's equations:
$$\small{E(\partial_\rho V_{\mu\nu})=dE(V_{\mu\nu})=dF=ddV=\begin{pmatrix}
    \partial_\ell \alpha^\ell-\omega_\ell(\epsilon_{\ell mn}\partial^m\alpha^n+\partial_0\beta_\ell) & i\partial_\ell \beta^\ell-i\omega_\ell(\epsilon_{\ell mn}\partial^m\beta^n-\partial_0\alpha_\ell) \\
    -i\partial_\ell \beta^\ell+i\omega_\ell(\epsilon_{\ell mn}\partial^m\beta^n-\partial_0\alpha_\ell) & -\partial_\ell \alpha^\ell+\omega_\ell(\epsilon_{\ell mn}\partial^m\alpha^n+\partial_0\beta_\ell)
\end{pmatrix}}$$
$$\partial_\ell \alpha^\ell=0 \ \ \Rightarrow \ \ \vec{\nabla}\cdot\vec{\alpha}=0$$
$$-\epsilon_{\ell mn}\partial^m\alpha^n-\partial_0\beta_\ell=0 \ \ \Rightarrow \ \ \vec{\nabla}\times\vec{\alpha}=-\partial_0\vec{\beta}$$
$$-\partial_\ell \beta^\ell=0 \ \ \Rightarrow \ \ \vec{\nabla}\cdot\vec{\beta}=0$$
$$\epsilon_{\ell mn}\partial^m\beta^n-\partial_0\alpha_\ell=0 \ \ \Rightarrow \ \ \vec{\nabla}\times\vec{\beta}=\partial_0\vec{\alpha}$$
\section{Magnetic Moments of the Leptons and Quarks} 

\subsection{Leptons}
Consider the kinetic Lagrangian for a weak isospin doublet composed of left chiral first generation leptons $L=\begin{pmatrix}
    \nu_L \\
    e_L
\end{pmatrix}$,
$$\Lag_L=iL^\dagger\tilde\Sigma^\mu(\partial_\mu-\frac{ig}{2}x_\ell X^\ell_\mu+\frac{ig'}{2}yY_\mu)L-L^\dagger mL $$
where the Lie algebra values $x_\ell=\{x_1=\begin{pmatrix}
0 & 1 \\
1 & 0 
\end{pmatrix} ,  x_2=\begin{pmatrix}
0 & -i \\
i & 0
\end{pmatrix} , x_3=\begin{pmatrix}
1 & 0 \\
0 & -1
\end{pmatrix}\}$ are weak isospin, $y=\begin{pmatrix}
    1 & 0 \\
    0 &1
\end{pmatrix}$ is weak hypercharge, and $m=\begin{pmatrix}
    m_\nu & 0 \\
    0 & m_e
\end{pmatrix}$ is the mass matrix. The equation of motion for the doublet is $$\frac{\partial\Lag_L}{\partial L^\dagger}=\tilde\Sigma^\mu(i\partial_\mu+\frac{g}{2}x_\ell X^\ell_\mu-\frac{g'}{2}yY_\mu)L-mL=0$$
which is unpacked into two equations coupling the two components
$$\begin{cases}
       \left[i\tilde\Sigma^\mu \partial_\mu+\frac{1}{2}\tilde\Sigma^\mu(gX^3_\mu-g'Y_\mu)-m_\nu\right]\nu_L 
+\frac{g}{2}\tilde\Sigma^\mu(X^1_\mu-iX^2_\mu)e_L=0 \\
       \left[i\tilde\Sigma^\mu\partial_\mu-\frac{1}{2}\tilde\Sigma^\mu(gX^3_\mu+g'Y_\mu)-m_e\right]e_L 
+\frac{g}{2}\tilde\Sigma^\mu(X^1_\mu+iX^2_\mu)\nu_L=0
    \end{cases} \ \ \ .$$
Inserting a planewave ansatz for each doublet component $e_L=\psi e^{-i(p_{e_L})_\mu x^\mu}$ and  $\nu_L=\phi e^{-i(p_{\nu_L})_\mu x^\mu}$ where $(p_{f_L})_\mu$ is the four-momentum of the left chiral component of each fermion $f=e,\nu$, as well as the charged weak boson definitions $W^\pm_\mu=\frac{1}{\sqrt{2}}(X^1_\mu\mp X^2_\mu)$ and inverting their $U(1)$-phase, we get 
$$\begin{cases}
       \left[\tilde\Sigma^\mu (p_{\nu_L})_{\mu}+\frac{1}{2}\tilde\Sigma^\mu(gX^3_\mu-g'Y_\mu)-m_\nu\right]\nu_L -\frac{g}{\sqrt{2}}\tilde\Sigma^\mu W^+_\mu e_L=0 \\
       \left[\tilde\Sigma^\mu (p_{e_L})_\mu-\frac{1}{2}\tilde\Sigma^\mu(gX^3_\mu+g'Y_\mu)-m_e\right]e_L 
-\frac{g}{\sqrt{2}}\tilde\Sigma^\mu W^-_\mu\nu_L=0
    \end{cases} \ \ .$$
Neglecting the neutrino terms, we obtain an equation for the left chiral electron
$$\left[\tilde\Sigma^\mu (p_{e_L})_\mu-\frac{1}{2}\tilde\Sigma^\mu(gX^3_\mu+g'Y_\mu)-m_e\right]e_L 
=0.$$
Using the definitions $X^3_\mu=\sin\theta A_\mu+\cos\theta Z_\mu$ and $Y_\mu=\cos\theta A_\mu-\sin\theta Z_\mu$ where $\theta\approx \sin^{-1}(\sqrt{.2396})= .511504$ is the mixing angle related to the weak coupling constants $g$ and $g'$ by $\tan^{-1}\theta=\frac{g'}{g}$, as well as the boson masses $m_Z=\frac{v\sqrt{g^2+g'^2}}{2}\approx91.19\ \text{GeV}$, $m_W=\frac{gv}{2}\approx80.36 \ \text{GeV}$, electric charge $e=g'\cos\theta$, and the Higgs vacuum expectation value $v\approx246\ \text{GeV}$ we rewrite the equation as
$$\left[\left(E_e-eA_0-\frac{m_Z}{v}\cos(2\theta)Z_0\right)-\tilde\Sigma^\ell \left(p_\ell-eA_\ell-\frac{m_Z}{v}\cos(2\theta)Z_\ell\right)-m_e\right]e_L 
=0.$$
Defining gauge-covariant energy and momentum eigenvalues $E_c\equiv(E_e-eA_0-\frac{m_Z}{v}\cos(2\theta)Z_0)$ and  $p_c\equiv\tilde\Sigma^\ell p^c_\ell= \tilde\Sigma^\ell(p_\ell-eA_\ell-\frac{m_Z}{v}\cos(2\theta)Z_\ell)$, we have
$$\left[E_c-p_c-m_e\right]e_L =0.$$
Taking relativistic kinetic energy as $T=E_c-m_e$, the equation is $(T-p_c)e_L=0$. Squaring the linear operator, $(T-p_c)$, gives the equation
$$\Rightarrow (T-p_c)^2e_L=0$$
$$\Rightarrow (T^2+p_c^2-2p_cT)e_L=0$$
$$\Rightarrow(T(E_c+m_e)-2m_eE_c+p_c^2-2p_cT)e_L=0$$
Noting that $p_c^2=(\tilde\Sigma^\ell p_{c\ell})^2=p_{c\ell} p_c^{\ell}-\omega_\ell\epsilon^{\ell mn}p_{cm}p_{cn}$, we take the non-relativistic limit $T\rightarrow 0$,
$$\Rightarrow E_c=\frac{p_c^2}{2m},$$
which yields the Pauli-Schrodinger equation
$$E_ce_L=\frac{p_c^2}{2m}e_L\Rightarrow E_ce_L=\frac{p_{c\ell} p_c^{\ell}-\omega_\ell\epsilon^{\ell mn}p_{cm}p_{cn}}{2m_e}e_L.$$
In full,

$$\small{(E_e-eA_0-\frac{m_Z}{v}\cos(2\theta)Z_0)e_L=\frac{(p_\ell-eA_\ell-\frac{m_Z}{v}\cos(2\theta)Z_\ell)^2}{2m_e}-\frac{e}{2m_e}\tilde\Sigma^\ell B_\ell-\frac{m_Z}{2m_ev}\cos(2\theta)\tilde\Sigma^\ell\zeta_\ell,}$$

where $E_\ell=\partial_0 A_\ell-\partial_\ell A_0$, $B_\ell=\epsilon_{\ell mn}\partial^mA^n$, $\phi_\ell=\partial_0 Z_\ell-\partial_\ell Z_0$, $\zeta_\ell=\epsilon_{\ell mn}\partial^mZ^n$, are the electric and magnetic fields corresponding to each neutral vector boson. Upon acquiring in the non-relativistic limit the magnetic moments of the electron with the photon and Z boson, we cannot help but feel as though we left one out. Looking at the original Lagrangian for the doublet, we see gauge-type couplings to the W boson as well, and the only difference between the W terms and the neutral terms is that they are off-diagonal. We naively assume that the electron has an equivalent notion of magnetic moment with the W boson. Seeing as the neutral moments are simply coupling constants divided by twice the mass, by analogy we assume that the charged moment follows the same pattern. Defining the W boson's derivatives as $\varepsilon^\pm_\ell=\partial_0 W^\pm_\ell-\partial_\ell W^\pm_0$ and $\beta^\pm_\ell=\epsilon_{\ell mn}\partial^mW^{\pm n}$, we contemplate the following expression for the energy of the charged magnetic moment
$${\mu_{\beta^+}}^\ell\beta^+_\ell=-\frac{\sqrt{2}m_W}{2m_ev}\tilde\Sigma^\ell\beta^+_\ell.$$
We infer from this expression that the electron has an opposing moment with the $W^-$, however its reasonable to assume these two particles are seldom near each other spatially (although near in time during decays) because they have the same electric charge. To define some useful terminology, just as we refer to the coupling strength of a particle to the photon as \textbf{electric charge}, we will refer in all cases to the coupling constant associated to a particle's interaction with the Z boson as \textbf{weak charge}, and the coupling constant associated to the $W^+$ as \textbf{electroweak charge} (the last of which is always inverted when referring to electroweak charge with the $W^-$). In the same way, we will refer to the moments of a particle with the magnetic fields of each of these bosons as electric/photonic, weak, and electroweak magnetic moments.

We can similarly contemplate two (or three) neutrino moments,
$${\mu_{\zeta-\nu}}^\ell\zeta_\ell=+\frac{1}{2m_\nu}\frac{m_Z}{v}\tilde\Sigma^\ell\zeta_\ell \ \ \   \text{and} \ \ \ {\mu_{\beta^--\nu}}^\ell\beta^-_\ell=-\frac{1}{2m_\nu}\frac{\sqrt{2}m_W}{v}\tilde\Sigma^\ell\beta^-_\ell.$$
which are very large compared to other particles because the neutrino mass (if it even exists) is very small. It is interesting that in spite of the neutrino being void of electric charge, it has a moment with a charged magnetic field.

\subsection{Quarks}
The up and down quark are also an isospin doublet $L_q=\begin{pmatrix}
    u_L \\
    d_L
\end{pmatrix}$, and each are additionally coupled to gluon fields $G^a_\mu$ for $a=1,...,8$
$$\Lag=iL_q^\dagger\tilde\Sigma^\mu(\partial_\mu-\frac{ig}{2}x_\ell X^\ell_\mu-\frac{ig}{6}yY_\mu-\frac{ic}{2}\lambda_aG^a_\mu)L_q-m_qL_q^\dagger L_q. $$
The equations of motion are
$$\begin{cases}
       \left[\tilde\Sigma^\mu p^{u}_{\mu}+\tilde\Sigma^\mu(\frac{1}{2}gX^3_\mu+\frac{1}{6}g'Y_\mu-\frac{1}{2}c\lambda_aG^a_\mu)-m_u\right]u_L -\frac{g}{\sqrt{2}}\tilde\Sigma^\mu W^+_\mu d_L=0 \\
       \left[\tilde\Sigma^\mu p^{d}_{\mu}+\tilde\Sigma^\mu(-\frac{1}{2}gX^3_\mu+\frac{1}{6}g'Y_\mu-\frac{1}{2}c\lambda_aG^a_\mu)-m_d\right]d_L -\frac{g}{\sqrt{2}}\tilde\Sigma^\mu W^-_\mu u_L=0
    \end{cases} \ \ \ .$$
The non-relativistic magnetic moments of the up quark are
\begin{align*}
    &\mu^\ell_{B-u}B_\ell=+\frac{1}{2m_u}\frac{2}{3}e\tilde\Sigma^\ell B_\ell,&\mu^\ell_{\zeta-u}\zeta_\ell=+\frac{1}{2m_u}\frac{m_Z}{v}(\cos^2\theta-\frac{1}{3}\sin^2\theta) \tilde\Sigma^\ell\zeta_\ell,\\ &\mu^\ell_{\beta^--u}\beta^-_\ell=-\frac{1}{2m_u}\frac{\sqrt{2}m_W}{v}\tilde\Sigma^\ell\beta^-_\ell,&\text{ and } \hspace{0.75in}\mu^\ell_{C^a-u}C^a_\ell=-\frac{c\lambda_a}{4m_u}\tilde\Sigma^\ell C^a_\ell,
\end{align*}

 and those of of the down quark are
 \begin{align*}
     &\mu^\ell_{B-d}B_\ell=-\frac{1}{2m_d}\frac{1}{3}e\tilde\Sigma^\ell B_\ell, &\mu^\ell_{\zeta-d}\zeta_\ell=-\frac{1}{2m_d}\frac{m_Z}{v}(\cos^2\theta+\frac{1}{3}\sin^2\theta)\tilde\Sigma^\ell\zeta_\ell,\\ &\mu^\ell_{\beta^+-d}\beta^+_\ell=-\frac{1}{2m_d}\frac{\sqrt{2}m_W}{v}\tilde\Sigma^\ell\beta^+_\ell, &\text{and}\hspace{0.75in}\mu^\ell_{C^a-d}C^a_\ell=-\frac{c\lambda_a}{4m_d}\tilde\Sigma^\ell C^a_\ell.
 \end{align*}

where $C^a_\ell$ is the magnetic field of the $a^{th}$ gluon.

\section{Magnetic Moments of Charged Bosons}

\subsection{The Charged Vector Potential}
It is known that the abelian kinetic term of the $W^\pm$ boson can be combined with certain terms in the gauge three-point self interaction (WWV) sub-Lagrangian to yield a gauge covariant kinetic term of the form
$$\Lag_{W^\pm-gauge}=-\frac{1}{2}X^+_{\mu\nu}X^{-\mu\nu}+m_W^2W^+_\mu W^{-\mu}$$
where $X^+_{\mu\nu}=D^*_\mu W^+_\nu-D^*_\nu W^+_\mu $ , $X^-_{\mu\nu}=D_\mu W^-_\nu-D_\nu W^-_\mu $ , with $D_\mu=\partial_\mu+iN_\mu$ where $N_\mu =e A_\mu+e\cot\theta Z_\mu$. 
The equation of motion for the four-vector $W^{-\mu}$ is
$$\Rightarrow \ \ \  D^*_\beta X^{+\beta\alpha}+m^2W^{+\alpha}=0.$$
Define $\varepsilon^{+}_{c \ \ell}\equiv-D^*_0 W^+_\ell-D_\ell W^+_0$ and $\beta^{+}_{c\ \ell}\equiv\epsilon_{\ell mn}D^{m}W^{+n}=R^m D_m W^+_\ell $  to be the gauge-covariant electric and magnetic fields of $W^{+\mu}$ and define $\eta_\ell\equiv\epsilon_{\ell mn}\partial^mN^n=e B_\ell+e\cot\theta\zeta_\ell$ a linear combination of the photon and Z-boson magnetic fields $B_\ell=\epsilon_{\ell mn}\partial^mA^n=R^m \partial_m A_\ell $ and $\zeta_\ell=\epsilon_{\ell mn}\partial^mZ^n=R^m \partial_m Z_\ell$. The components of $R_\ell$ are $\mathfrak{so}(3)$ generators: $$R_1=\begin{pmatrix}
    0 & 0 & 0 \\
    0 & 0 & -1 \\
    0 & 1 & 0
\end{pmatrix} , \ \ R_2=\begin{pmatrix}
    0 & 0 & 1 \\
    0 & 0 & 0 \\
    -1 & 0 & 0
\end{pmatrix} , \ \ R_3=\begin{pmatrix}
    0 & -1 & 0 \\
    1 & 0 & 0 \\
    0 & 0 & 0
\end{pmatrix}.$$
Embedding relevant objects in the spin representation:
$$\mathcal{D}\equiv\Gamma^\mu D^*_\mu=\begin{pmatrix}
    \partial_0-iN_0 & i\omega_\ell(\partial_\ell+iN_\ell) \\
    -i\omega_\ell(\partial_\ell+iN_\ell) & -\partial_0+iN_0
\end{pmatrix} \equiv \begin{pmatrix}
    D^*_0 & i\omega_\ell D_\ell \\
    -i\omega_\ell D_\ell & -D^*_0
\end{pmatrix}$$
$$W^+\equiv\Gamma^\mu W^+_\mu=\begin{pmatrix}
    W^+_0 & -i\omega_\ell W^+_\ell \\
    i\omega_\ell W^+_\ell & -W^+_0
\end{pmatrix}$$
$$\Gamma^\mu\Gamma^\nu X^+_{\mu\nu}=\begin{pmatrix}
    -\omega_\ell\beta^{+}_{c \ \ell} & i\omega_\ell \varepsilon^{+}_{c \ \ell} \\
    i\omega_\ell \varepsilon^{+}_{c \ \ell} & -\omega_\ell\beta^{+}_{c \ \ell}
\end{pmatrix}=\frac{1}{2}[\mathcal{D},W^+] $$
If we define a gauge-covariant Lorenz condition, $D^*_\mu W^{+\mu}=0$, the gauge covariant field strength can be expressed in terms of the embedded objects $\mathcal{D} \ ,\  W^+$ as $\Gamma^\mu\Gamma^\nu X^+_{\mu\nu}=\mathcal{D}W^+$.  

The equation of motion for $W^{-\mu}$ can then be written as
$$\mathcal{D}^2W^++m_W^2W^+=0,$$
which we might call the gauge-covariant Klein-Gordon equation, or equivalently, the gauge-covariant Proca-Maxwell's (GCPM) equations:
$$ \mathcal{D}^2W^++m^2W^+=0 \ \ \Rightarrow \ \ \begin{cases}
      \vec{D}\cdot\vec{\varepsilon}_c^++m^2W^+_0=0 \\
      \vec{D}\times\vec{\varepsilon}_c^+=-D^*_0\vec{\beta}_c^+ \\
      \vec{D}\cdot\vec\beta_c^+=0 \\
      \vec{D}\times\vec\beta_c^++m^2\vec{W}^+=D^*_0\vec{\varepsilon}_c^+
    \end{cases}$$
Plugging into Ampere's law the definitions of $\vec\varepsilon_c^+$ and $\vec\beta_c^+$,  we obtain a wave equation for the charged vector potential that is homogeneous when the scalar potential $W^+_0=0$
$$\left[{D^*_0}^2-(\vec{D}\cdot\vec{D})+(iR^\ell \eta_{\ell})+m^2\right]\vec{W}^+=0.$$
Defining the plane-wave eigenvalues $\vec{D}W^{+\mu}=ip_cW^{+\mu}$ and $D^*_0W^{+\mu}=-iE_cW^{+\mu}$, the equation is
$$\left[-E_c^2+p_c^{2}+(iR^\ell \eta_{\ell})+m^2\right]\vec{W}^+=0.$$
Define relativistic kinetic energy $T=E_c-m$.
$$\Rightarrow \ \ \ \left[-T(E_c+m)+p_c^{2}+(iR^\ell \eta_\ell)\right]\vec{W}^+=0$$
In the non-relativistic limit, $E_c\approx m$ $\ \Rightarrow \ \ T_{NR}=\frac{1}{2m}\left[p_c^{2}+(iR^\ell \eta_\ell) \right]$.
The Proca-Maxwell-Pauli-Schrodinger (PMPS) equation for $\vec{W}^+$ is $E_c\vec{W}^+=T_{NR}\vec{W}^+$:
$$(\partial_0-iN_0)\vec{W}^+=\left[\frac{-1}{2m}(\partial_\ell+iN_\ell)^2+\frac{1}{2m}(iR^\ell \eta_\ell) \right]\vec{W}^+.$$
In full,
$$\small{(\partial_0-i(eA_0+e\cot\theta Z_0))\vec{W}^+=\left[\frac{-1}{2m_W}(\partial_\ell+i(eA_\ell+e\cot\theta Z_\ell))^2+\frac{1}{2m_W}(+iR^\ell (eB_\ell+e\cot\theta\zeta_\ell)) \right]\vec{W}^+}.$$
The charged vector potential $\vec{W}^+$ has both electric and weak charge (which are both positive) and thus has both photonic and weak magnetic moments $\vec\mu_W=\frac{1}{2m_W}(+iR^\ell (eB_\ell+e\cot\theta\zeta_\ell))$.
 
The PMPS equation for the conjugate vector potential $\vec{W}^-$ is the conjugate equation
$$(\partial_0+iN_0)\vec{W}^-=\left[\frac{-1}{2m_W}(\partial_\ell-iN_\ell)^2+\frac{1}{2m_W}(-iR^\ell \eta_\ell) \right]\vec{W}^-$$
It is interesting to note that the U(1) Lie algebra in the gauge covariant derivative forced the real antisymmetric matrices $R^\ell$ we began with to be Hermitian. Unitary gauge coupling forced the birth representation to transform the charged vector field as the fundamental representation.

\subsection{The Charged Magnetic Field}
We obtained the GCPM equations via a spin representation of the spacetime algebra whose generators transform as an $\mathfrak{su}(2)\oplus\mathfrak{su}(2)$ representation of the Lorentz group. It is now useful to employ a representation of $Cl_{(1,3)}\mathbb{R}$ whose generators are an $\mathfrak{so}(3)\oplus\mathfrak{so}(3)$ representation
$$\kappa^\mu=\left(\kappa^0= \begin{pmatrix}
    I_{3\times3} & 0 \\
    0 & -I_{3\times3}
\end{pmatrix} , \kappa^\ell=\begin{pmatrix}
    0 & R_\ell \\
    R_\ell & 0
\end{pmatrix}\right)$$
built from the $R_\ell\in\mathfrak{so}(3)$. Defining $\Psi^+=\begin{pmatrix}
    \vec\beta_c^+ \\
    \vec\varepsilon_c^+
\end{pmatrix}$, we can rewrite the GCPM Faraday's and Ampere's equations as
$$\kappa^\mu D^*_\mu\Psi^++\frac{1}{2}(I_{6\times6}-\kappa^0)m^2\vec{W}^+=0$$
or
$$(R^\ell D_\ell)\vec{\varepsilon}_c^+=-D^*_0\vec{\beta}_c^+$$
$$(R^\ell D_\ell)\vec{\beta}_c^++m^2\vec{W}^+=D^*_0\vec{\varepsilon}_c^+$$
Performing elimination gives
$$(R^\ell D_\ell)(R^\ell D_\ell)\vec{\beta}_c^++m^2(R^\ell D_\ell)\vec{W}^+=-{D^*_0}^2\vec\beta_c^+$$
$$\Rightarrow \ \ {D^*_0}^2\vec\beta_c^+-(\vec{D}\cdot\vec{D})\vec{\beta}_c^++(iR^\ell \eta_{\ell})\vec\beta_c^++m^2(R^\ell D_\ell)\vec{W}^+=0.$$
Recognizing $(R^\ell D_\ell)\vec{W}^+=\vec{D}\times\vec{W}^+=\epsilon_{\ell mn}D^{m}W^{+n}\equiv\vec{\beta}^+ $ as the positively charged magnetic field (not the gauge-covariant version), the equation is
$$\left[(\partial_0-iN_0)^2-(\partial_\ell+iN_\ell)^2+(iR^\ell \eta_\ell)\right]\vec\beta_c^+ +m^2\vec\beta^+=0.$$
The magnetic moments of the charged magnetic field $\vec\beta^+$ are the same as it's vector potential's: $\vec\mu_B=\frac{e}{2m_W}(i\vec{R})$ and $\vec\mu_\zeta=\frac{e}{2m_W}\cot\theta(i\vec{R})$. 

\section{A Model of Dynamical Parity Violation}

Taking the non-relativistic limit of various equations of motion serves it's purpose in identifying the magnetic moments of given particles. However, the non-relativistic limit (NRL) conceals a few key features of these general magnetic interactions. The main problem is that in taking the NRL, we have to square the gauge-covariant momentum. This will always result in the same spin-orientation of coupling (either left or right) for particles with the same sign of electric and/or weak and/or electroweak charge. In other words, taking the NRL forces the impression that an electron rotating left-handedly is to be polarized by an external field identically to a right-handed electron. This does not make sense if a particle's magnetic moment is a vector-valued quantity. The direction of a particle's spin and the sign of it's quantum numbers determine it's intrinsic magnetic moments and thus determine the direction of said moments' coupling with any external field.

Define a "natural" or "chiral" frame for viewing a fermion to be any frame in which it's helicity is the same as it's intrinsic chirality. Then, in a natural frame colinear with a propagating left chiral electron, it's intrinsic photonic, weak, and electroweak magnetic moments each point in the direction of its translational velocity because the electron has negative electric, weak, and electroweak charge. In a natural frame colinear with a propagating right-chiral electron, the poles are opposite.

Now, in a natural frame that is non-colinear, if we consider the moving electron to be a current source for photonic, weak, and electroweak magnetic fields (not the intrinsic field but the Amperian field that appears around a wire with current flowing through it), there are no differences between the right and left chiral electrons from the Amperian perspective because their charges are identical. There is always a frame where both the moments and the Amperian fields are of comparable strength.

The key idea then lies in the fact that a particle's intrinsic magnetic moment is a vector valued quantity, while the Amperian magnetic field of any flavor is a pseudovector quantity. Combining this perspective with the magnetic moments previously derived, we realize that the electroweak magnetic moment of the electron is to be influenced by the neutral Amperian fields with which the charged magnetic field has moments. From here it's easy to see that the vector value of a particle's electroweak magnetic moment implies that the influence will be opposite under parity transformation. This effect can be visualized classically - FIG. 1 and FIG. 2.
\begin{figure}[ht]
    \centering
    \includegraphics[width=0.25\linewidth]{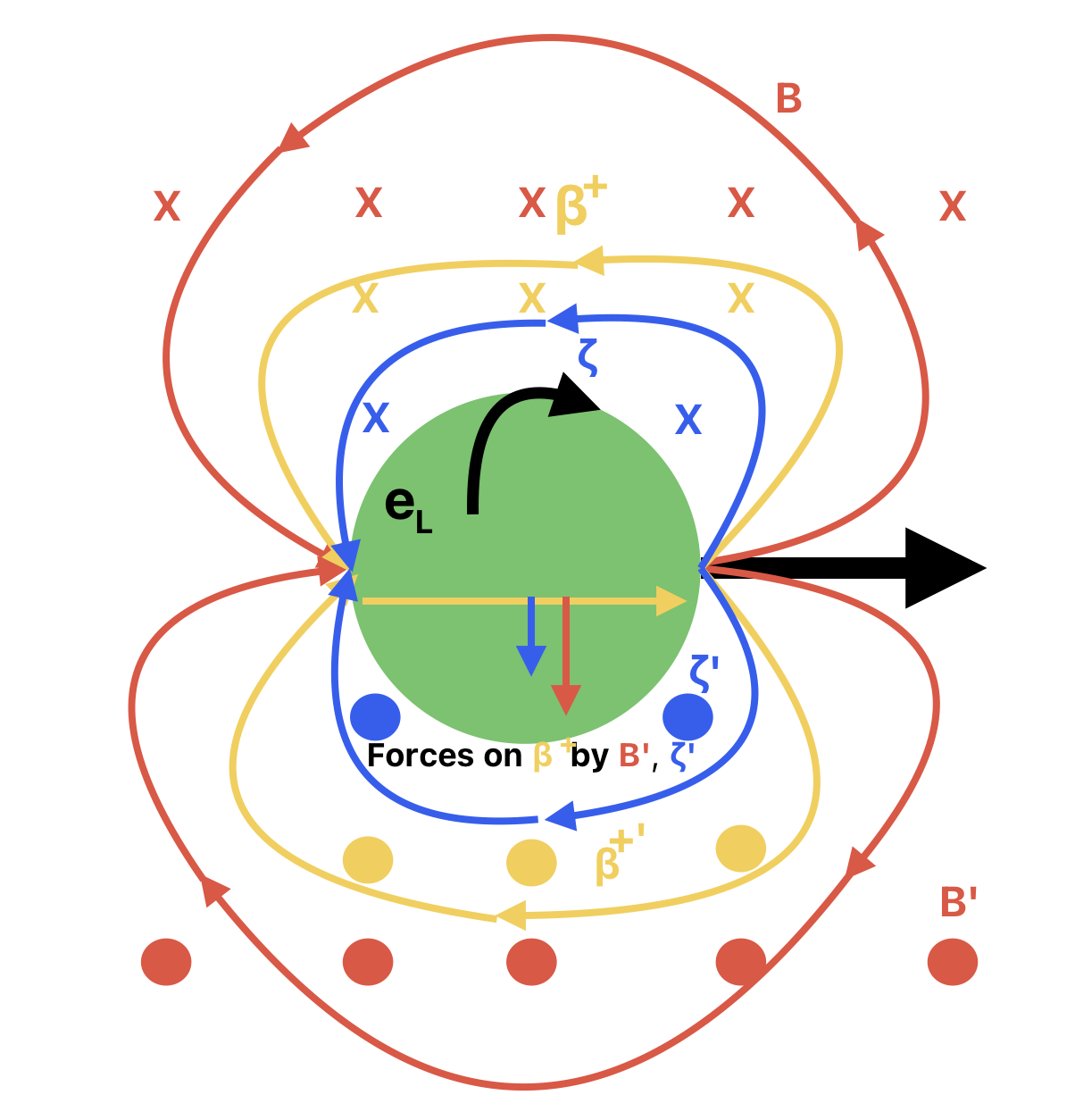}
    \caption{A classical depiction of the left chiral electron with it's weak, electroweak, and photonic magnetic moments in increasing order outward according to length scales. The primed fields are the Amperian fields. A $\beta^+$ field line is depicted just below the axis of rotation with two arrows showing the direction of force exerted by each neutral Amperian field.}
    \label{fig:Classical Left Chiral Electron}
\end{figure}

Concerns that arise about the conservation of charge when looking at FIG. 1 are quelled once one realizes that there are $\beta^-$ field lines flowing simultaneously and oppositely to the $\beta^+$ lines shown.

\begin{figure}[ht]
    \centering
    \includegraphics[width=0.25\linewidth]{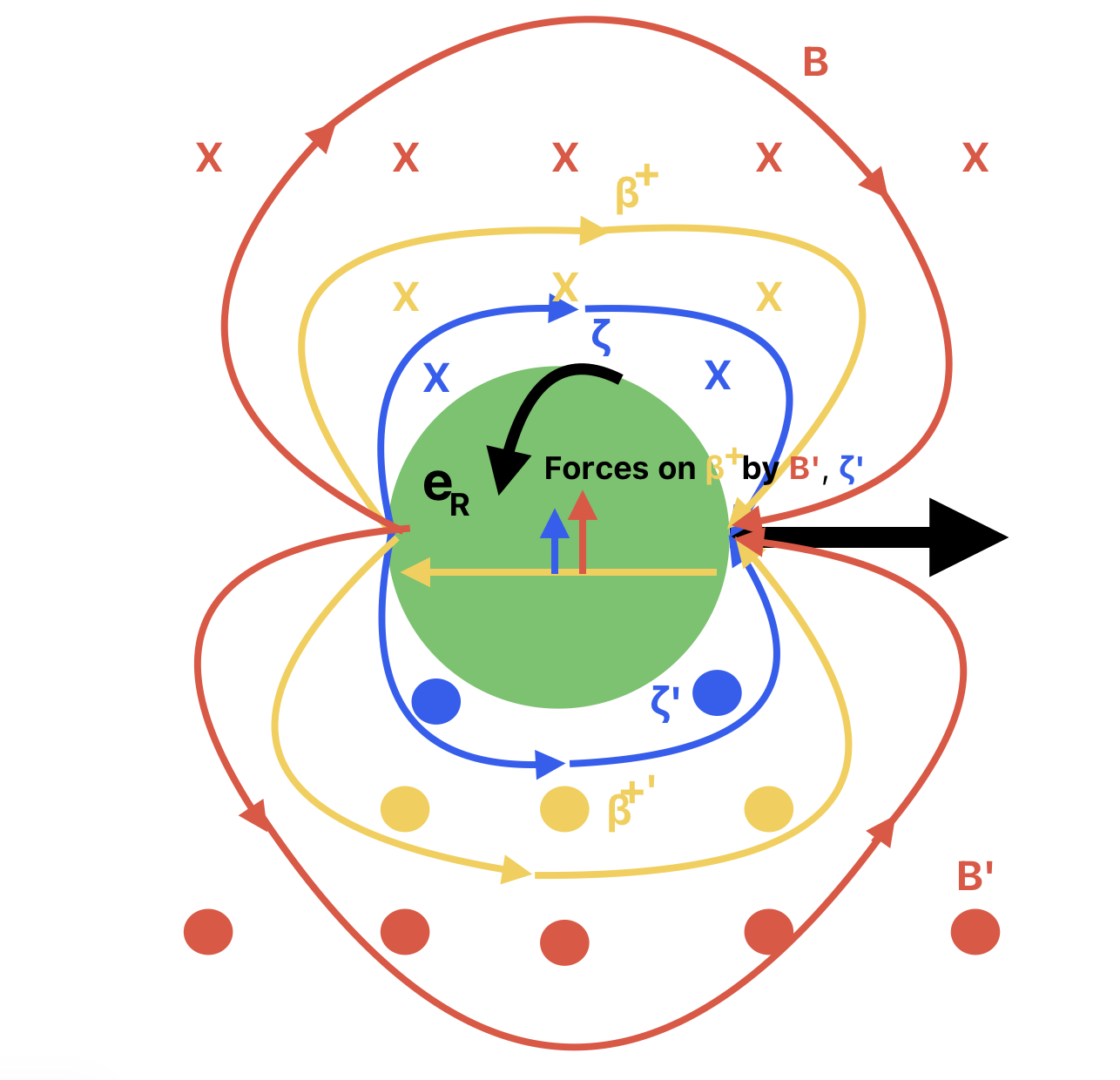}
    \caption{The right chiral electron depicted with magnetic moments opposite to that of the left chiral counterpart, but with the same Amperian fields.}
    \label{fig: Classical Right Chiral Electron}
\end{figure}

The charged magnetic moments of the electron are either pushed outward or inward by the neutral Amperian fields, and which one it is depends on it's chirality. If an electron's ability to interact with other particles via a specific boson is at all determined by the "reach" of it's corresponding magnetic moment, this effect would contribute to the chiral asymmetry of charged weak interactions. Perhaps the right chiral fermions are coupled to weak isospin before spontaneous symmetry breaking (SSB) after all, but due to the neutral bosons that emerge under conditions of SSB, their ability to interact via the charged bosons are heavily suppressed. It could be that nature favors the left-chiral particles for participation in these interactions simply because it is easier to get them involved than their right-chiral counterparts. 

That being said, maybe there are right chiral neutrinos among us. The same diagram drawn for a left chiral neutrino - which uniquely among the fermions carry only weak and electroweak quantum numbers - also matches experimental trends - FIG. 3.

\begin{figure}[ht]
    \centering
    \includegraphics[width=0.25\linewidth]{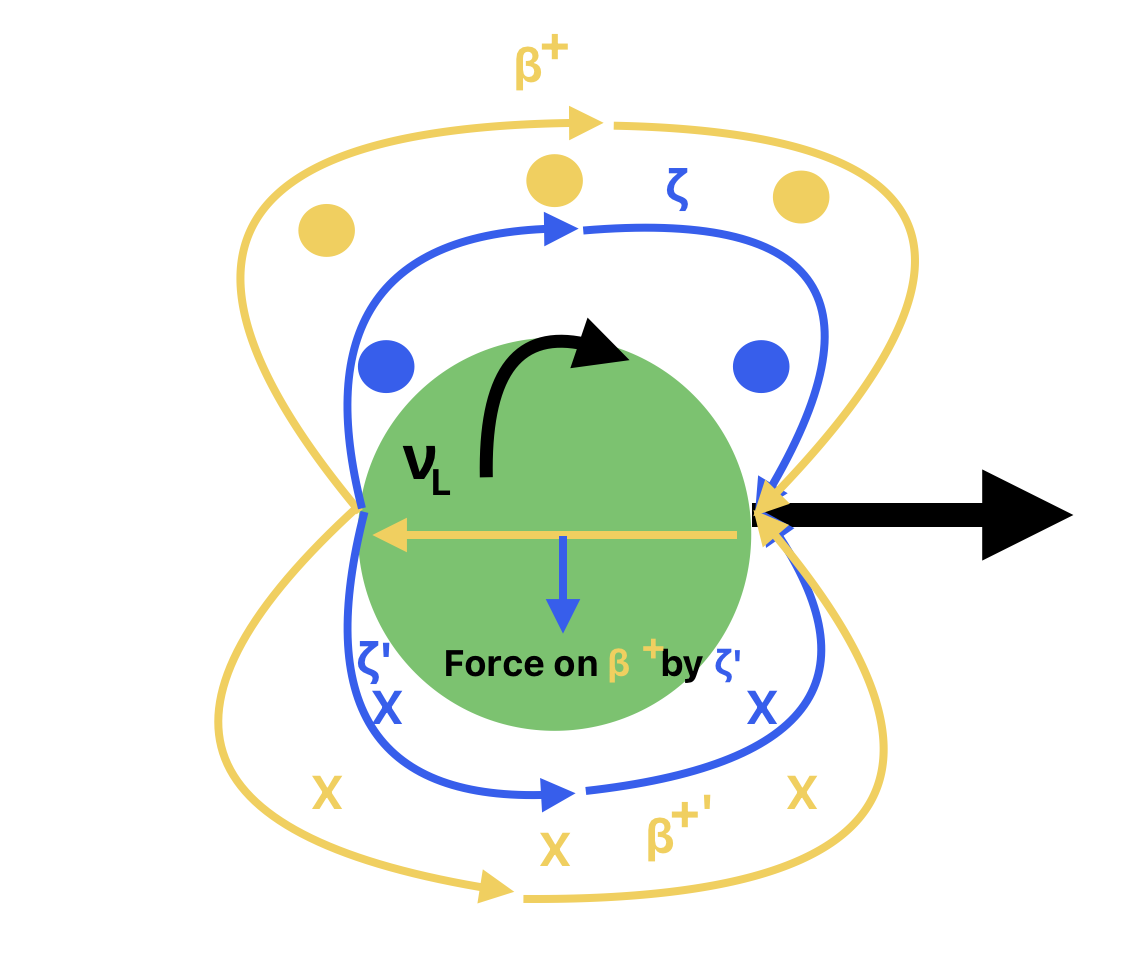}
    \caption{The left chiral neutrino depicted with the outward torque exerted by the Z boson on its $\beta^+$ moment.}
    \label{fig: Classical Left Chiral Neutrino}
\end{figure}

Since the neutrino has positive weak and electroweak charge, we see that the Amperian weak field produced by it's translational motion should aid a left chiral neutrino's participation in interactions involving the charged bosons whilst suppressing that of a right chiral neutrino. Historically interesting discussions on chirality and parity in particle physics can be found in \cite{Landau1957,Zeldovich1958}.

\section{Discussion, Experiments, and Conclusions}

The above diagrams supply great physical insight despite the fact that fundamental particles have never been observed to have finite geometry of any kind. We are inclined to believe that this reasoning remains valid in the case of quantum fields. The claim that the electron's moment with the $W^\pm$ boson can be considered independently of the presence of a neutrino remains questionable. In spite of this and other concerns, we are able to deduce that certain chiral components of fermions could be more available to interact via charged weak bosons given how their charged and neutral moments interact with their Amperian fields. These charged magnetic fields are likely to be mostly vacuum fluctuations at energies below the mass of the $W$ boson and only observable within very small proximity of the electron itself. Presuming that electroweak moments and/or Amperian fields can align and enhance their overall strength, a beam of left chiral electrons should leave a stronger electroweak signature than a beam of right chiral electrons. If such a charged Amperian field exists around a beam of left chiral electrons, it could allow for the observation of right chiral neutrinos, as left chiral neutrinos moving near the beam should be polarized and scattered by the beam more readily than right chiral ones. 

There is concern surrounding charged (heavily self-interacting) magnetic fields existing macroscopically. If a quantum perspective is taken, however, it is imaginable that a proper theoretical (likely nonlinear) treatment of a self-interaction driven instability present in electroweak force fields may improve our current understanding of weakly unstable phenomena. If nuclei have overall electroweak moments due to an aligning effect, we expect those of radioactive nuclei to be of greater strength than those of stable ones. At the level of nuclear constituents, the presence of such fields might contribute to the instability of the neutron relative to the proton. Much more speculatively, they may reveal information about the mechanisms behind flavor change. Perhaps an alignment of quark or lepton electroweak moments somehow allows for the exchange of charges, masses, and ultimately flavors. It is unclear at the moment how important the roles of gluon magnetic moments are in understanding processes involving quarks. The exact implications of such magnetic fields existing at the fundamental level warrant further investigation.

An examination of data collected from various charged weak interactions, or high energy beams of electrons such as those at LEP, could reveal a route to detecting the electroweak moments of electrons, neutrinos, nuclei, etc. To state the obvious, any means of detection with current technology is likely to be indirect and ultimately reliant on electromagnetism. Rydberg atoms may be a possible tool for detecting these exotic magnetic moments~\cite{DeStefano2024,Behary2025}. It is natural to suspect a connection to CP violating processes such as Kaon decay, but at the moment this connection is unclear.

In conclusion, any fermionic particle in possession of a quantum number associated with an interaction mediated by a vector boson possesses a magnetic moment with that vector boson. The electrons have three (or four), the neutrinos have two (or three), and the quarks have eleven (or twelve) moments. It remains to be determined how to properly account for the proposed neutral magnetic suppression of the charged fermion moments at the Lagrangian level. In any case, these concepts merit proper transference into the landscape of perturbative calculations. 

\section*{Acknowledgments}
 
The authors are grateful to Duncan Weathers for his valuable discussions, comments, and suggestions. JHA thanks Jaskeerat (Jace) Singh, Hamidreza (Hamid) Khajoei, and Aaron Richardson for numerous invigorating conversations.

\bibliography{ReferencesMu,RefQuaternion}

\end{document}